\documentclass{article}
\begin{document}
\begin{center}{\Huge{\bf Pr\'ecis: Quantum Feedback}}\end{center}

\begin{center}{H. M. Wiseman}\end{center}


\newcommand{\beq}{\begin{equation}}
\newcommand{\eeq}{\end{equation}}
\newcommand{\dg}{^\dagger}
\newcommand{\smallfrac}[2]{\mbox{$\frac{#1}{#2}$}}
\newcommand{\half}{\smallfrac{1}{2}}

\setcounter{footnote}{0}

\begin{quote}
The following is the body of page ix of the PhD  thesis {\em Quantum Trajectories and 
Feedback} by H.M. Wiseman (Physics Department, University of Queensland, 1994), which is downloadable as a postscript file at \verb$http://www.sct.gu.edu.au/~sctwiseh/PhDThesis.ps.z$. It is (as it describes itself) a very brief technical summary of the most important results therein. \\
\end{quote}

Consider an optical field in one dimension with canonical commutation relations
\beq
[b(z,t),b\dg(z',t)]=v\delta(z-z'),
\eeq
propagating at speed $v$. Let it be coupled to a system at 
position $z=0$ by the Hamiltonian
\beq
V_{\rm dipole}(t) = {\rm i}[b\dg(0,t)c(t) - c\dg(t)b(0,t)],
\eeq
where a rotating-wave approximation has been used \footnote{C. W. Gardiner,
{\em Quantum Noise} (Springer-Verlag, Berlin, 1991).}. Let the output 
photocurrent be fed back to control the system via
\beq
V_{\rm feedback}(t) = Z(t) b\dg(v\tau,t)b(v\tau,t).
\eeq
For a vacuum input field [zero eigenstate of $b(z,t)$ for $z<0$], 
the explicit increment in an arbitrary system operator is
\begin{eqnarray}
ds &=& {\rm i}[H,s]dt - [s,c\dg] \left( \half c+ 
b_0 \right) dt+ \left(\half c\dg + b_0\dg  \right) 
[s,c]dt \nonumber \\
&& + \; [ c\dg(t-\tau) + b_0\dg(t-\tau)] 
\left( e^{{\rm i}Z}s e^{-{\rm i}Z} - s\right)[ c(t-\tau) + b_0(t-\tau)]dt ,
\end{eqnarray}
where $b_0(t) \equiv b(0^-,t)$ commutes with $s(t)$ and obeys
\beq
b_0(t)b_0\dg(t)dt = 1,
\eeq
with other such moments vanishing ${}^2$. For finite delay $\tau$, 
the output field operator $b_0(t-\tau)+c(t-\tau)$ commutes with $s(t)$, 
but for $\tau=0$ it does not. In the latter case, one obtains
the corresponding feedback master equation
\beq
\dot{\rho}=-{\rm i}[H,\rho] + e^{-{\rm i}Z}c\rho c\dg e^{{\rm i}Z}
 - \half\{ c\dg c, \rho \}.
\eeq
This has an obvious interpretation in terms of the quantum jumps 
associated with photodetection in the theory of quantum trajectories 
\footnote{H. J. Carmichael, {\em An Open Systems Approach to Quantum Optics}
(Springer-Verlag, Berlin, 1993)}.

\end{document}